\documentclass[aps,prl,showpacs,twocolumn,groupedaddress,
nobibnotes]{revtex4-2}

\usepackage[colorlinks,
          linkcolor=blue,
            citecolor=blue,
            urlcolor=blue
           ]{hyperref}
\usepackage{amsmath,amsthm,amssymb}
\usepackage{array}
\usepackage{graphicx}
\usepackage{fancyhdr}
\usepackage{color}
\usepackage{extarrows}
\usepackage{enumerate}
\usepackage{epstopdf}
\usepackage{bm}
\usepackage{comment}
\usepackage{float}
\usepackage{ulem}
\usepackage{booktabs}
\begin{document}

\newcommand{\wl}[1]{{\color[rgb]{.9,.1,.1}{#1}}}
\newcommand{\wlc}[1]{{\color[rgb]{.9,.1,.8}{[Zhou: {\it #1}\,]}}}
\newcommand{\wlx}[1]{{\color[rgb]{.5,.5,.5}{\sout{#1}}}}

\newcommand{\han}[1]{{\color[rgb]{.5,.1,.5}{#1}}}
\newcommand{\hanc}[1]{{\color[rgb]{.6,.2,.8}{[HL: {\it #1}\,]}}}
\newcommand{\hanx}[1]{{\color[rgb]{.5,.1,.8}{\sout{#1}}}}
\newcommand{\B}[1]{{\color[rgb]{.0,.0,1.0}{#1}}}

\title{$5/9-$Magnetization Plateau and Spin Supersolidity in YCu$_3$(OD)$_{7-x}$Br$_{2+x}$\\ under Magnetic Fields up to 120~T}

\author{Hiroaki~Hayashi$^{2}$}
\thanks{These authors contributed equally to this work.}
\author{Han~Li$^{3}$}
\thanks{These authors contributed equally to this work.}
\author{Feng-Feng~Song$^{2}$}
\thanks{These authors contributed equally to this work.}
\author{Xu-Guang~Zhou$^{1,2,\B{*}}$}
\thanks{Corresponding authors: xgzhou@hmfl.ac.cn; w.li@itp.ac.cn; ymatsuda@issp.u-tokyo.ac.jp}

\author{Akira~Matsuo$^{2}$}
\author{Taeyun~Kim$^{4}$}
\author{Enze~Lv$^{5}$}
\author{Yuto~Ishii$^{2}$}
\author{Zhe~Qu$^{1}$}
\author{Gang~Su$^{5}$}
\author{Koichi~Kindo$^{2}$}
\author{Kwang-Yong~Choi$^{4}$}
\author{Wei~Li$^{5}$}
\thanks{Corresponding authors: xgzhou@hmfl.ac.cn; w.li@itp.ac.cn; ymatsuda@issp.u-tokyo.ac.jp}
\author{Yasuhiro~H.~Matsuda$^{2}$}
\thanks{Corresponding authors: xgzhou@hmfl.ac.cn; w.li@itp.ac.cn; ymatsuda@issp.u-tokyo.ac.jp}

\affiliation{$^1$Anhui Key Laboratory of Low-Energy Quantum Materials and Devices, High Magnetic Field Laboratory, Hefei Institutes of Physical Science, Chinese Academy of Sciences, Hefei 230031, China}

\affiliation{$^2$Institute for Solid State Physics, University of Tokyo, Kashiwa, Chiba 277-8581, Japan}

\affiliation{$^3$School of Physical Science and Engineering, Beijing Jiaotong University, Beijing 100044, China}

\affiliation{$^4$Department of Physics, Sungkyunkwan University, Suwon, Republic of Korea}

\affiliation{$^5$Institute of Theoretical Physics, Chinese Academy of Sciences, Beijing 100190, China}


\begin{abstract}
We performed high-precision magnetization measurements up to 120~T on three compositions of the newly discovered kagome antiferromagnet YCu$_3$(OD)$_{7-x}$Br$_{2+x}$ (YCOB), revealing a previously unobserved 5/9 fractional magnetization plateau. All YCOB samples with different Br$^-$ concentrations exhibit nearly identical magnetization curves below 60~T, whereas the 5/9 plateau appears at markedly different fields in the ultrahigh-field regime. By modeling the experimental data using tensor-network calculations, we derive the effective spin Hamiltonians for the YCOB family with three spatially anisotropic Heisenberg couplings (the 3$J$-type model), which quantitatively reproduces the measured magnetization processes and captures the composition-dependent evolution of the 5/9 plateau. Furthermore, our theoretical analysis suggests the emergence of a spin supersolid phase in the field window between the 1/3 and 5/9 plateaus, which is sensitive to spin exchange parameters and accounts for the significant variation in the critical fields of the 5/9 plateau observed among different YCOB compositions.
\end{abstract}
\maketitle

\noindent\textit{Introduction.---}
Field-induced exotic phases in frustrated quantum magnets provide a fertile platform for exploring strongly correlated systems and collective spin behaviors~\cite{Balents2010, Mila2015, Kitaev2006}. These phases include quantum spin liquids (QSLs)~\cite{Kasahara2018}, fractional magnetization plateaus~\cite{Shangguan2023}, spin superfluid via magnon Bose-Einstein condensation (BEC) phases~\cite{Nomura2026, Zhou2020}, and even spin supersolid states~\cite{Xiang2024, Shu2026}. Such exotic quantum spin states hold potential for diverse applications, including topological quantum computation~\cite{Kitaev2006} and ultralow-temperature refrigeration~\cite{Li2024}. Over the past decades, a variety of candidate materials have been identified as approximate realizations of the corresponding theoretical models, such as Shastry-Sutherland lattice compound SrCu$_2$(BO$_3$)$_2$~\cite{Matsuda2013, Kageyama1999}, Kitaev material $\alpha$-RuCl$_3$~\cite{Kasahara2018}, and triangular-lattice XXZ Heisenberg compound Na$_2$BaCo(PO$_4$)$_2$~\cite{Xiang2024}, among many others.

As one of the most prominent examples of frustrated spin systems, the kagome-lattice Heisenberg antiferromagnet (KHAF)~\cite{Ran2007, Depenbrock2012, Yan2011, ZhouYi2017} has long inspired the search for material realizations. Early efforts centered on herbertsmithite~\cite{Shores2005, Fu2015, Mendels2010, Mendels2007, Khuntia2020, Pilon2013}, yet Zn/Cu site mixing prevented a definitive interpretation of its magnetic behavior~\cite{Lee2007, Vries2008, Mendels2010, Kermarrec2014, Wang2021}. Through systematic chemical substitutions and structural refinements~\cite{Sun2016, Zorko2019, Puphal2017}, researchers synthesized YCu$_3$(OD)$_{7-x}$Br$_{2+x}$ (YCOB), which crystallizes in the high-symmetry $P\bar{3}m1$ space group and stands as the closest known realization of the ideal KHAF~\cite{Chen2020}. Subsequent experiments have accumulated mounting evidence for a proximate QSL ground state~\cite{Lee2024, Liu2022, Shivaram2024, Xu2024, Lishiliang2025, Zeng2024, Lu2022, Zeng2022, Suetsugu2025}, reinforcing the view that YCOB lies very close to the ideal KHAF limit. Nevertheless, recent studies report residual OH$^-$/Br$^-$ site disorder that may induce bond-dependent 3$J$-type interactions and randomness~\cite{Liu2022, Lu2022, Lishuo2024}, raising questions about how faithfully this compound represents the pristine KHAF model.

High magnetic fields provide a powerful probe of spin states and their transitions, offering direct insight into microscopic spin interactions. For the ideal KHAF model, theory predicts a series of fractional magnetization plateaus at 1/9, 1/3, 5/9, and 7/9 of the saturated magnetization~\cite{Nishimoto2013}, with a supersolid phase between the 1/3 and 5/9 plateaus~\cite{Plat2018}. Experimentally, the 1/9 plateau --- absent in herbertsmithite and many other candidate materials~\cite{Okuma2019, Kato2024} --- has recently been observed in YCOB under high magnetic fields~\cite{Jeon2024, Zheng2025pnas, Suetsugu2024}. This plateau is particularly intriguing because it has been proposed to host a field-induced QSL state stabilized by strong quantum fluctuations~\cite{Nishimoto2013, Jeon2024, Zheng2025pnas, Suetsugu2024, Zheng2025prx, Morita2024jsps,
He2024prl,
He2024prb}. YCOB also exhibits the characteristic 1/3 plateau predicted for the ideal KHAF model~\cite{Suetsugu2024,He2025}. Whether YCOB hosts higher fractional plateaus and additional exotic field-induced phases remains an open question , calling for further ultrahigh-field magnetization studies.

In this work, we report the observation of the 5/9 fractional magnetization plateau in a series of YCOB compounds ($x \simeq 0.05$, 0.50, and 0.95) via high-precision magnetization measurements up to 120~T. While the 1/9 and 1/3 plateaus are nearly identical across compositions $x$, the 5/9 plateau exhibits pronounced composition dependence, with its onset field falling well below the $\sim$130~T predicted for the ideal KHAF. By fitting the experimental data with DMRG and iPEPS calculations, we extract an effective 3$J$-type spin model with spatially anisotropic exchange, which could quantitatively reproduce the magnetization curves of YCOB compounds. We further propose a spin supersolid phase between the 1/3 and 5/9 plateaus whose width is tuned by exchange anisotropy, naturally explaining the composition-dependent 5/9-plateau critical fields. Our work establishes the microscopic spin Hamiltonian for YCOB and lays a foundation for exploring exotic quantum spin states in this intriguing family of kagome compounds.
\\

\noindent\textit{High-field measurements and data fittings.---}
Magnetization measurements were performed on single crystals of YCOB with compositions $x \simeq 0.05$, 0.50, and 0.95. Pulsed magnetic fields were generated using the single-turn coil (STC) technique, reaching up to 120~T in a vertical-type configuration~\cite{Miura2003research, Matsuda2026}. 
The magnetization is measured using a parallel compensated pickup coil, as described in Refs.~\cite{Takeyama2012, Matsuda2013, Zhou2020, Zhou2023possible, Zhou2025}. 
To achieve sufficient energy for generating magnetic fields up to 120~T with a 14~$\phi$ single-turn coil, two sets of capacitor banks with total energy of 200~kJ are employed instead of one, marking the first attempt to perform magnetization measurements under such extreme conditions. This configuration provides a more homogeneous magnetic field  than previous 12~$\phi$ setup~\cite{Matsuda2013} and allows the use of a larger-diameter parallel-compensated pickup coil~\cite{Zhou2023possible,Zhou2025}.
For all measurements, the consistency between the up- and down-sweep magnetization curves has been carefully checked. All experiments were conducted at an initial temperature of 2~K under the magnetic field along the $c$ axis.
The experimental magnetization curves are analyzed using DMRG and iPEPS simulations. Further details of the numerical calculations are provided in the Supplemental Materials (SM)~\cite{SM}.\\

\begin{figure}[tbp]
\begin{center}
\includegraphics[width = 1.0\linewidth]{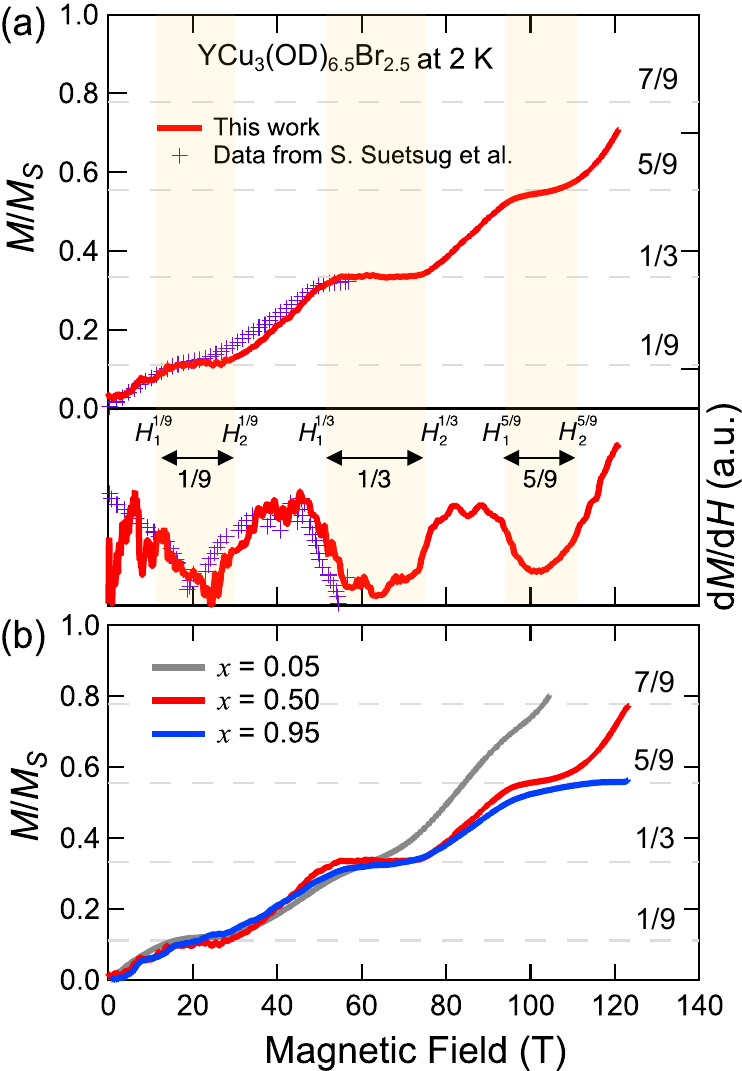}
\caption{(a) Magnetization process and $\mathrm{d}M/\mathrm{d}H$ data of YCOB ($x \simeq 0.50$) measured at 2~K. Previous measurements up to 60~T~\cite{Suetsugu2024} are denoted as purple symbols as a comparison, where a very good agreement is seen to the present measurements. The observed magnetization plateaus are labeled by the yellow shadow and black arrows. (b) Magnetization process for different compositions of YCOB ($x \simeq 0.05$, 0.50, and 0.95) measured at 2~K.}
\label{MH}
\end{center}
\end{figure}

\noindent\textit{Observation of 5/9 magnetization plateau with composition dependence.---}
Figure~\ref{MH}(a) shows the magnetization process and the corresponding $\mathrm{d}M/\mathrm{d}H$ curves of YCOB ($x \simeq 0.50$) measured up to 120~T (red curve). Our results demonstrate excellent agreement with the nondestructive pulsed-field data from Ref.~\cite{Suetsugu2024}, shown for comparison. Three fractional magnetization plateaus are clearly identified at 1/9, 1/3, and 5/9 of the saturation magnetization $M_S$, as indicated by the black arrows.
For convenience, we define the critical fields of the $n/9$ plateaus as $H^{n/9}_{1,2}$, where the subscripts $1$ and $2$ represent the onset and end critical fields of each plateau, respectively. As indicated by the onset and end points of the arrows in Fig.~\ref{MH}(a), the critical fields are determined as: $H^{1/9}_\mathrm{1} \simeq 13$~T, $H^{1/9}_\mathrm{2} \simeq 26$~T, $H^{1/3}_\mathrm{1} \simeq 52$~T, $H^{1/3}_\mathrm{2}\simeq 73$~T, $H^{5/9}_\mathrm{1}\simeq 96$~T and $H^{5/9}_\mathrm{2}\simeq 108$~T, respectively.

\begin{figure*}[tbp]
\begin{center}
\includegraphics[width = 1.0\linewidth]{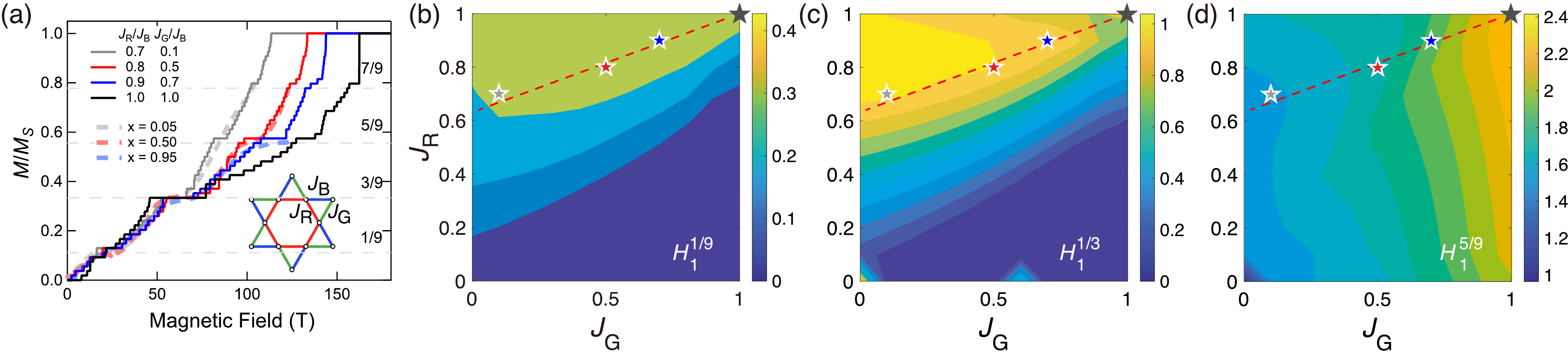}
\caption{(a) Experimental and DMRG results for the normalized magnetization curves of YCOB. The gray, red, and blue solid (dashed) curves represent the calculated (measured) data for Br compositions $x \simeq 0.05$, $0.50$, and $0.95$, respectively. The black curve corresponds to the ideal KHAF model. The exchange parameters, given in units of $J_{\rm B}$, are the best-fitting values obtained from the analysis described in the Supplemental Materials~\cite{SM}. The inset illustrates the definition of the three exchange couplings $J_{\rm R}$, $J_{\rm G}$, and $J_{\rm B}$ on the nonequivalent bonds.
(b-d) Contour plots of fields $H_1^{n/9}$ for the $n/9$ ($n=1,3,5$) plateaus calculated with DMRG. The best-fitting parameters for three materials and the ideal KHAF parameter set are indicated by the stars in all panels. The results in (b-d) are in natural unit with $J_{\rm B} = 1$, which corresponds to 6.46~meV for $x \simeq 0.05$, 6.89~meV for $x \simeq 0.50$, and 6.81~meV for $x \simeq 0.95$. The contour interval is set to $\Delta H =0.1$ across panels (b-d).
}
\label{PD}
\end{center}
\end{figure*}

We extended the ultrahigh-field magnetization measurements to the compositional extremes of the YCOB family, $x \simeq 0.05$ and $x \simeq 0.95$. The full set of magnetization curves for all three compositions is compared in Fig.~\ref{MH}(b), with the extracted critical fields listed in Table~SI of the SM~\cite{SM}. The STC data were calibrated against non-destructive pulsed-field measurements, as detailed in the End Matter.

Strikingly, below 60~T the three curves are nearly indistinguishable, indicating that the low-field magnetic response is insensitive to Br$^-$ content. Above this threshold, however, the behavior diverges markedly. For $x \simeq 0.05$, the 1/3 plateau develops a finite slope, in contrast to the flat plateau seen at $x \simeq 0.50$, and the 5/9 plateau is barely discernible, with magnetization rising steeply toward saturation. In the opposite limit, $x \simeq 0.95$, the 1/3 plateau closely matches that of $x \simeq 0.50$, whereas the 5/9 plateau extends to fields exceeding 120~T, the maximum accessible field in the present experiment.

This minimal compositional sensitivity at low fields naturally explains why earlier studies --- which were confined to fields below 60~T --- failed to resolve the influence of Br$^-$ concentration on the magnetization process: the dramatic differences emerge only in the ultrahigh-field regime.

\noindent\textit{Effective 3$J$-type KHAF model for YCOB.---}
In Fig.~\ref{PD}(a), the black curve shows the calculated magnetization of the ideal KHAF model (see details in SM~\cite{SM}), with the energy scale fixed by fitting the previously observed 1/9 and 1/3 plateaus of YCOB ($x\simeq0.50$). On this scale, the ideal KHAF model predicts the $H^{5/9}_1 \simeq 130$~T~\cite{Suetsugu2024, Jeon2024, Nishimoto2013}, yet the experimentally observed critical fields lie substantially lower. For YCOB ($x \simeq 0.50$), for instance, the experimental onset field is reduced by roughly 30~T. This discrepancy points to a departure from the ideal kagome limit. Previous DFT calculations and $M$-$T$ analyses indeed suggest that OH$^-$/Br$^-$ site mixing in YCOB ($x \simeq 0.50$) induces slight lattice distortions~\cite{Liu2022, Lu2022}, which render three inequivalent nearest-neighbor superexchange couplings on each kagome triangle [see inset of Fig.~\ref{PD}(a)]. 

Accordingly,, we describe the system by a 3$J$-type Heisenberg model with bond-dependent couplings $J_{\rm R}$, $J_{\rm G}$, and $J_{\rm B}$, as illustrated in the inset of Fig.~\ref{PD}(a). At this stage, we do not consider the effects of randomness. Following Refs.~\cite{Morita2024, Liu2022, Lu2022}, we write the effective spin Hamiltonian as

\begin{equation}
\begin{aligned}
\hat{H}
&= \sum_{\langle i,j\rangle_{\mathrm{\gamma}}} J_{\mathrm{\gamma}}\, \mathbf{S}_i \cdot \mathbf{S}_j -\, g \mu_\mathrm{B} H \sum_i S_i^z \ ,
\end{aligned}
\label{eq:3-J_H}
\end{equation}

Here, $\langle i,j\rangle_{\mathrm{\gamma}}$ ($\mathrm{\gamma}$ = $\mathrm{R,G,B}$) labels the three inequivalent nearest-neighbor bonds within each kagome triangle, and the last term is the Zeeman coupling. With the g-factor fixed at $g\simeq 2.2$~\cite{Jeon2024}, we fit the experimental magnetization curves in the $(J_{\rm R},J_{\rm G},J_{\rm B})$ parameter space using numerical methods including DMRG~\cite{WhitePRL,Schollwoeck2011} and iPEPS~\cite{Jordan2008, Xie2014, Liu2015, Mei2017, Niu2022, Xu2023, Ferrari2023}.

By systematically varying $J_\mathrm{R}/J_\mathrm{B}$ and $J_\mathrm{G}/J_\mathrm{B}$ from 0 to 1 while keeping $J_\mathrm{B}=1$ as the reference energy scale, we minimize the mean absolute deviation $\mathcal{L} = \sum_{i=1}^N |M_{\rm Cal}^i - k M_{\rm Exp}^i| /N$ between the calculated and measured magnetization~\cite{SM}. Figure~\ref{PD}(a) shows that the optimal DMRG curves reproduce the experimental data with excellent accuracy. The best-fit parameters are summarized in Table~\ref{tab:params}.
\begin{table}[b]
\caption{\label{tab:params} Best-fit exchange parameters for the 3J Heisenberg model of YCOB. The energy scale $J_\mathrm{B}$ and the dimensionless ratios $J_\mathrm{R}/J_\mathrm{B}$, $J_\mathrm{G}/J_\mathrm{B}$ were determined by DMRG and iPEPS calculations with an estimated uncertainty of $\pm (0.05$--$0.1).$}.
\begin{ruledtabular}
\begin{tabular}{ccccc}
Composition & $J_\mathrm{R}/J_\mathrm{B}$ & $J_\mathrm{G}/J_\mathrm{B}$ & $J_\mathrm{B}$ (meV) & Fit quality\\
\hline
$x\simeq 0.05$ & $0.7$ & $0.1$ & $6.46$ & qualitative\\
$x\simeq 0.50$ & $0.8$ & $0.5$ & $6.89$ & quantitative\\
$x\simeq 0.95$ & $0.9$ & $0.7$ & $6.81$ & quantitative\\
\end{tabular}
\end{ruledtabular}
\end{table}
The 3$J$ model quantitatively captures the magnetization of the $x \simeq 0.50$ and $0.95$ samples. Notably, the energy scale for $x \simeq 0.50$, $J_\mathrm{B} \approx 80$~K (6.89~meV), agrees with prior estimates~\cite{Zeng2024}. For $x \simeq 0.05$, the agreement is only qualitative, as the finite 1/3-plateau slope and the suppression of the 5/9 plateau indicate additional effects --- most likely higher levels of disorder~\cite{Suetsugu2024} --- beyond the 3$J$ framework. The full loss landscape from both DMRG and iPEPS is included in End Matter.

The 3$J$ model also provides a unified explanation for the sharply contrasting low- and high-field behaviors. Figures~\ref{PD}(b)--(d) map the calculated onset fields $H_1^{n/9}$ of the $n/9$ ($n=1,3,5$) plateaus across the $(J_\mathrm{R}/J_\mathrm{B},J_\mathrm{G}/J_\mathrm{B})$ plane. The fitted parameters for all three compositions lie in a region where the $1/9$- and $1/3$-plateau onsets vary only weakly, explaining their nearly identical low-field magnetization. At higher fields, this homogeneity breaks down dramatically: the critical field $H_1^{5/9}$ of the $5/9$ plateau depends sensitively on exchange anisotropy, giving rise to the pronounced composition dependence observed in experiment. Linear spin-wave analysis corroborates these high-field trends, as detailed in the End Matter.

Across the YCOB family, the fitted exchange parameters reveal a systematic trend: increasing Br$^-$ concentration progressively suppresses the 3$J$ anisotropy, driving the system closer to --- though still appreciably away from --- the ideal KHAF limit. A plausible microscopic origin is the evolution of OH$^-$/Br$^-$ site disorder at the two crystallographically inequivalent Br$^-$ positions~\cite{Lu2022}: as the Br$^-$ content rises, the occupancy and disorder at these sites tends to equalize, giving rise to a more isotropic superexchange environment. This chemical tunability offers a realistic pathway for achieving the ideal KHAF, as illustrated in Figs.~\ref{PD}(b)--(d).\\

\begin{figure}[tbp]
\begin{center}
\includegraphics[width = 1\linewidth]{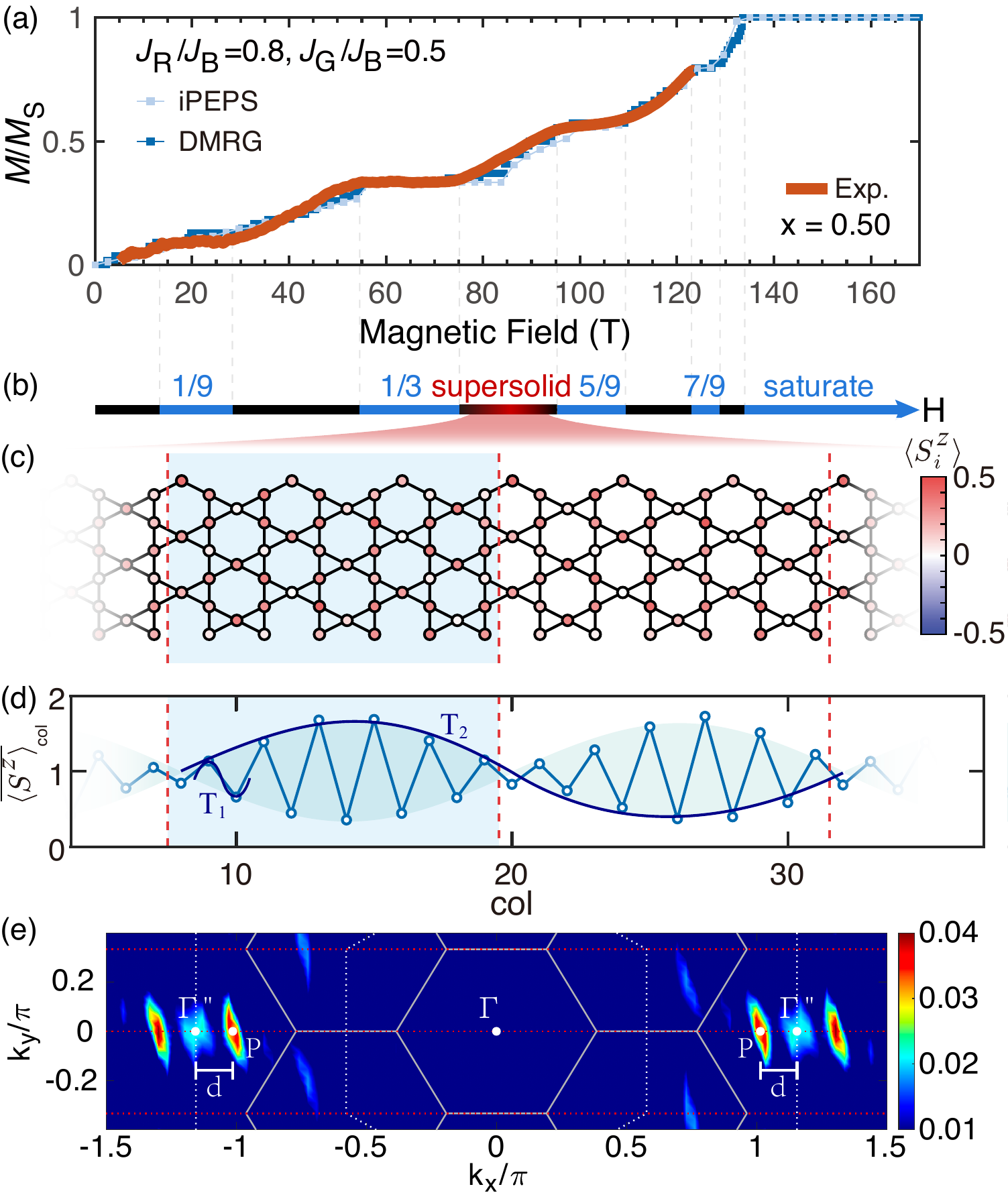}
\caption{(a) Many-body magnetization calculations for the YCOB material with $x \simeq 0.50$. The experimental data are shown as a red curve, the DMRG results are presented by the blue curve with the best-fit 3$J$ parameter set, and the iPEPS results with the same parameter are shown with light blue color. 
(b) Schematic field-induced phase diagram.
(c) Local moment $\langle S_{\rm i}^z \rangle$ on each site $i$ calculated with DMRG method, with parameter set $J_{\rm R} = 0.8$, $J_{\rm G}=0.5$, $J_{\rm B}=1$, and a fixed magnetization of $M/M_{\rm S} \simeq 4/9$. The color intensity of the circle  at each site $i$ indicates the magnitude and sign of the value, indicated by the color bar on the right. (d) The average of local moment $\overline{\langle S_z \rangle}_{\rm col}$ (see main text). A enlarged period is marked by the blue region. 
(e) The Fourier transform of $\langle S_{\rm i}^z \rangle$, $\Gamma''$ point is located at ${\textbf k_{\Gamma''}} = {\textbf b}_{\rm 1}/2 = [2\pi/\sqrt{3},\ 0]$ in the extended BZ. The peak is shifted to point P, with a distance $d = (T_{\rm 1}/T_{\rm 2})(2\pi/\sqrt{3})$ from the $\Gamma''$ point.
}
\label{Cal}
\end{center}
\end{figure}

\noindent\textit{Ultrahigh-field spin supersolid phase between 1/3 and 5/9 plateaus.---}
We now discuss the origin of the pronounced composition dependence of the 5/9 magnetization plateau in YCOB compounds and the presence of spin supersolid phase.
A recent theoretical study suggested the emergence of a supersolid phase between the 1/3 and 5/9 plateaus in both the ideal and the $J$--$J^{\prime}$ KHAF models~\cite{Plat2018}, where $J^{\prime}$ corresponds to $J_{\mathrm{G}} = J_{\mathrm{B}}$. 
In that framework, the supersolid phase gradually shrinks as the inter-hexagon coupling decreases. 
Consequently, the competition between inter- and intra-hexagon couplings within the supersolid phase may play an essential role in tuning the magnetization process between the 1/3 and 5/9 plateaus. 

To examine whether the supersolid phase persists within the 3$J$-type KHAF model and connects to experimental observations, we carried out DMRG calculations and focused on YCOB with $x \simeq 0.50$. This composition provides a robust 5/9 plateau, allowing for a reliable determination of the effective model parameters, as shown in Fig.~\ref{Cal}(a). Based on these parameters, we then examined the nature of the supersolid phase in the field window between the 1/3 and 5/9 plateaus. The resulting field-induced schematic phase diagram is shown in Fig.~\ref{Cal}(b). 
As illustrated in Figs.~\ref{Cal}(c) and (d), real-space distributions of the spin expectation values at each column, $\overline{\langle S_z \rangle}_{\rm col}$, reveal a periodic structure with a unit cell three times larger than that of the underlying Hamiltonian, indicating a spontaneous breaking of translational symmetry. 

Moreover, the Fourier transform of the local moment,
$S({\textbf k}) = \sum_{\rm i} \langle S_{\rm i}^z \rangle e^{i{\textbf k}\cdot{\textbf r_{\rm i}}}/N$,
reveals that the dominant peak in the Brillouin zone (BZ) is shifted away from the $\Gamma''$ point $(k_{\rm x}=2\pi/\sqrt{3},\ k_{\rm y}=0)$ to a new commensurate position P, as shown in Fig.~\ref{Cal}(e). This shift originates from the real-space beating pattern visible in the column-averaged magnetization $\overline{\langle S_z \rangle}_{\rm col}$ [Fig.~\ref{Cal}(d)]: a short period $T_{\rm 1}=2$ columns is modulated by a longer envelope $T_{\rm 2}=24$ columns. Writing the envelope as $A = $ $\sin(2\pi x/T_{\rm 2})\sin(2\pi x/T_{\rm 1})$ 
and after Fourier transform, the displacement of the peak position away from $\Gamma''$ point in the extended BZ is $d= (T_{\rm 1}/T_{\rm 2})(2\pi/\sqrt{3})$, in excellent agreement with the P-point observed in Fig.~\ref{Cal}(e).

Combined with the $U(1)$ spin-rotation-symmetry breaking between the magnetization plateaus~\cite{Nishimoto2013, Plat2018} (also see SM~\cite{SM}), these findings strongly indicate a spin supersolid phase in YCOB under ultrahigh magnetic fields. As the system moves away from the ideal KHAF limit, the supersolid region is progressively reduced, producing a steeper magnetization slope between the 1/3 and 5/9 plateaus. It is this sensitivity of the supersolid phase to exchange anisotropy that explains the pronounced composition-dependent high-field magnetization behavior across the YCOB family.

In the highly anisotropic $3J$ kagome model ($x\sim0.05$ in our case), the spin supersolid phase is suppressed 
as exchange anisotropy breaks the equivalence of hexagon clusters and disrupts coherent 
inter-hexagon coupling. Consequently, the mobility and phase coherence of defect magnons 
are strongly reduced. Once the spin supersolid phase is destabilized, the 1/3 and 5/9 magnetization 
plateaus collapse into compressible or inhomogeneous phases, as the excitation gap protecting 
the plateau becomes spatially nonuniform and ultimately vanishes on a global scale.\\

\noindent\textit{Conclusion and outlook.---}
In summary, we have performed magnetization measurements up to 120~T on YCOB compounds with three Br$^-$ concentrations ($x \simeq 0.05$, 0.50, and 0.95). While all samples exhibit nearly identical magnetization behavior below 60~T, their ultrahigh-field responses differ markedly. These contrasting behaviors are quantitatively captured by a spatially anisotropic 3$J$-type KHAF model with bond-dependent couplings $J_\mathrm{R}$, $J_\mathrm{G}$, and $J_\mathrm{B}$ (see Table~\ref{tab:params}). Our results identify the Br$^-$ concentration as a chemically tunable knob for modulating magnetic interactions, opening avenues for stabilizing QSLs in these KHAFs~\cite{Zeng2024, Lishiliang2025}. 


Furthermore, our 3$J$-type model analysis suggests the existence of a spin supersolid phase in the field window between the 1/3 and 5/9 plateaus. The pronounced composition dependence of the high-field magnetization can be attributed to the sensitivity of this supersolid phase to 3$J$ coupling anisotropy~\cite{Plat2018}. 
Future experiments using high-precision ultrasound or magnetostriction measurements, now feasible up to 80~T and potentially extendable to 140~T~\cite{Nomura2023}, could provide a direct thermodynamic probe of the predicted supersolid state.\\

\noindent\textit{Acknowledgments.---} 
X.-G.Z. and Z.Q. were supported by the National Key Research and Development Program of China (Grants No. 2024YFA1611101), the Strategic Priority Research Program of Chinese Academy of Sciences (Grant No. XDB1270102),  Anhui Provincial Major S \& T Project (Grant No. s202305a12020005), Anhui Provincial Natural Science Foundation (Grant No. 2408085J025). 
H.H. and Y.H.M. were funded by JSPS KAKENHI, Grant-in-Aid for Transformative Research Areas (A) No. 23H04859 and No.~23H04860, Grant-in-Aid for Scientific Research (B) No.~23H01117, and Grant-in-Aid for Challenging Research (Pioneering) No.~20K20521. 
F.-F.S. was supported by JSPS Grant-in-Aid for Early-Career Scientists (Grant No.~25K17311) and Grant-in-Aid for JSPS Fellows (Grant No.~25KF0183).
H.L. and W.L. were supported by the National Natural Science Foundation of China [Grants No.~12222412, No.~12447101 (W.L.), and No.~12404177 (H.L.)], and the Talent Fund of Beijing Jiaotong University [Grant No.~2025JBRC003 (H.L.)]. K.-Y.C was supported by the National
Research Foundation (NRF) of Korea (Grants No. RS-2023-
00209121 and No. 2020R1A5A1016518).

We thank Yuan Yao, Akihiko Ikeda, Yoshimitsu Kohama, Naoki Kawashima and Tsuyoshi Okubo for insightful discussions.
Y.H.M., H.H., Y.I., and X.-G.Z. especially thank Hironobu Sawabe for his technical assistance in maintaining laboratory safety and equipment.\\

\noindent\textit{Data availability.---} 
The data that support the findings of this
article are available from the authors upon reasonable request.


\sloppy
\bibliography{Kagome}

\section{End Matter}

\begin{figure}[h!]
    \begin{center}
    \includegraphics[width = 1\linewidth]{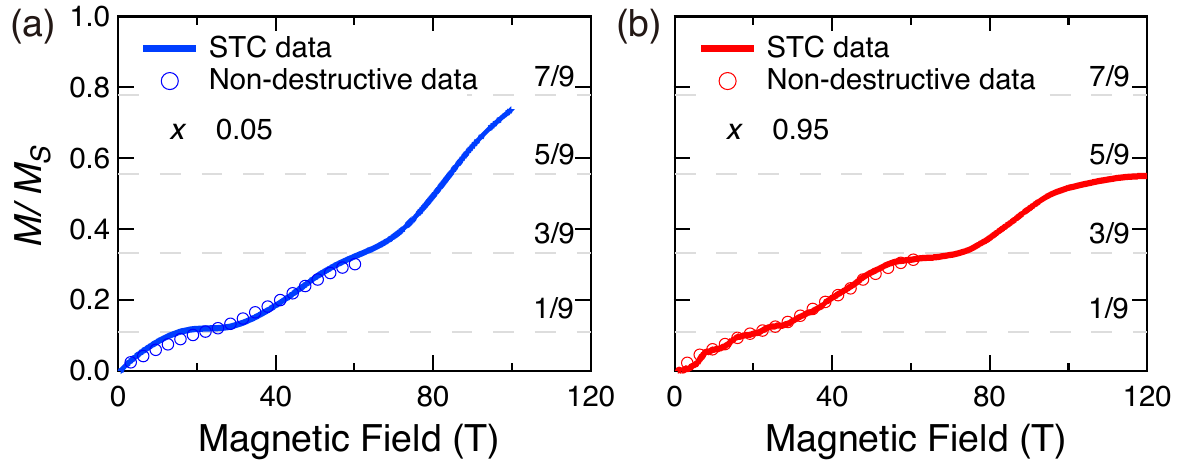}
     \caption{Low-field calibration of STC high-field magnetization data for YCOB with (a) $x \simeq 0.05$ and (b) $x \simeq 0.95$.}
     \label{Nondes}
    \end{center}
    \end{figure}

\noindent\textit{Experimental details.---}
Figure~\ref{Nondes} shows the low-field calibration of the magnetization data obtained from the STC high-field measurements. The low-field data were measured by the induction method, where pulsed magnetic fields up to 60~T were generated using a non-destructive magnet at a temperature of 1.3~K. For both YCOB compositions ($x \simeq 0.05$ and 0.95), the low-field results show good agreement with the high-field data.

\noindent\textit{Determination of the model parameters for YCOB.---}
Through fitting the experimentally measured $M$-$H$ curves for YCOB materials, 
we determine the effective model parameters 
corresponding to each compositions $x \simeq 0.05$, $x \simeq 0.50$, and $x \simeq 0.95$.
Since exchanging $J_{\rm B}$ and $J_{\rm G}$ should only correspond to a mirror reflection of the system 
and should yield identical magnetization curves, 
here we fix $J_{\rm B} = 1$ and vary $J_{\rm R}$ and $J_{\rm G}$ from 0 to 1 to define the parameter space for fitting.
For each set of parameters, we first adjust the energy scale $J_{\rm scale}$ = $J_{\rm B}$ to minimize the discrepancy 
between the computed magnetization ($M_{\rm Cal}$) curve and the experimentally measured one ($M_{\rm Exp}$), 
thereby calculating loss functions for each parameter set 
through 
\begin{equation}
    \mathcal{L} = \sum_{i=1}^N |M_{\rm Cal}^i - k M_{\rm Exp}^i| /N, \label{eq:fitting}
\end{equation}
where $k = J_{\rm scale} \times k_{\rm B} /g/\mu_{\rm B}$ and $g=2.2$ is the Land{\'e} factor.
This loss function measures the average absolute deviation between theory and experiment over all sampled field points. Minimizing $\mathcal{L}$ therefore selects the parameter set that best reproduces the overall magnetization process, including the plateau positions and slopes.

\begin{figure}[tbp!]
    \begin{center}
    \includegraphics[width = 1\linewidth]{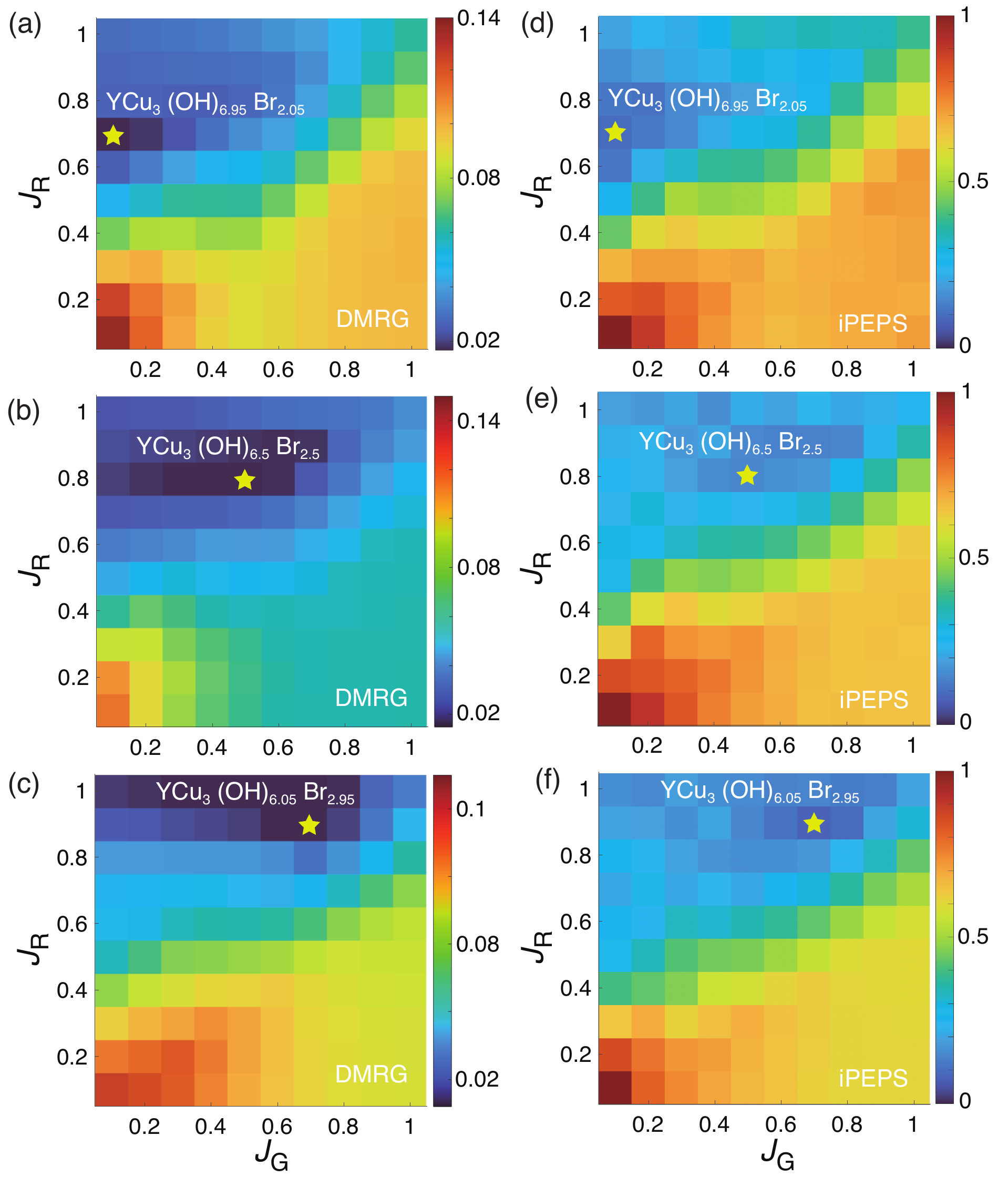}
     \caption{Loss map of the fitting results for YCOB materials 
     obtained from the (a-c) DMRG simulations and (d-f) iPEPS calculations.
     The best-fitting parameters are indicated by the yellow star in each panels.}
     \label{DMRGfittings}
    \end{center}
    \end{figure}

\begin{figure}[t!]
    \begin{center}
    \includegraphics[width = 0.9\linewidth]{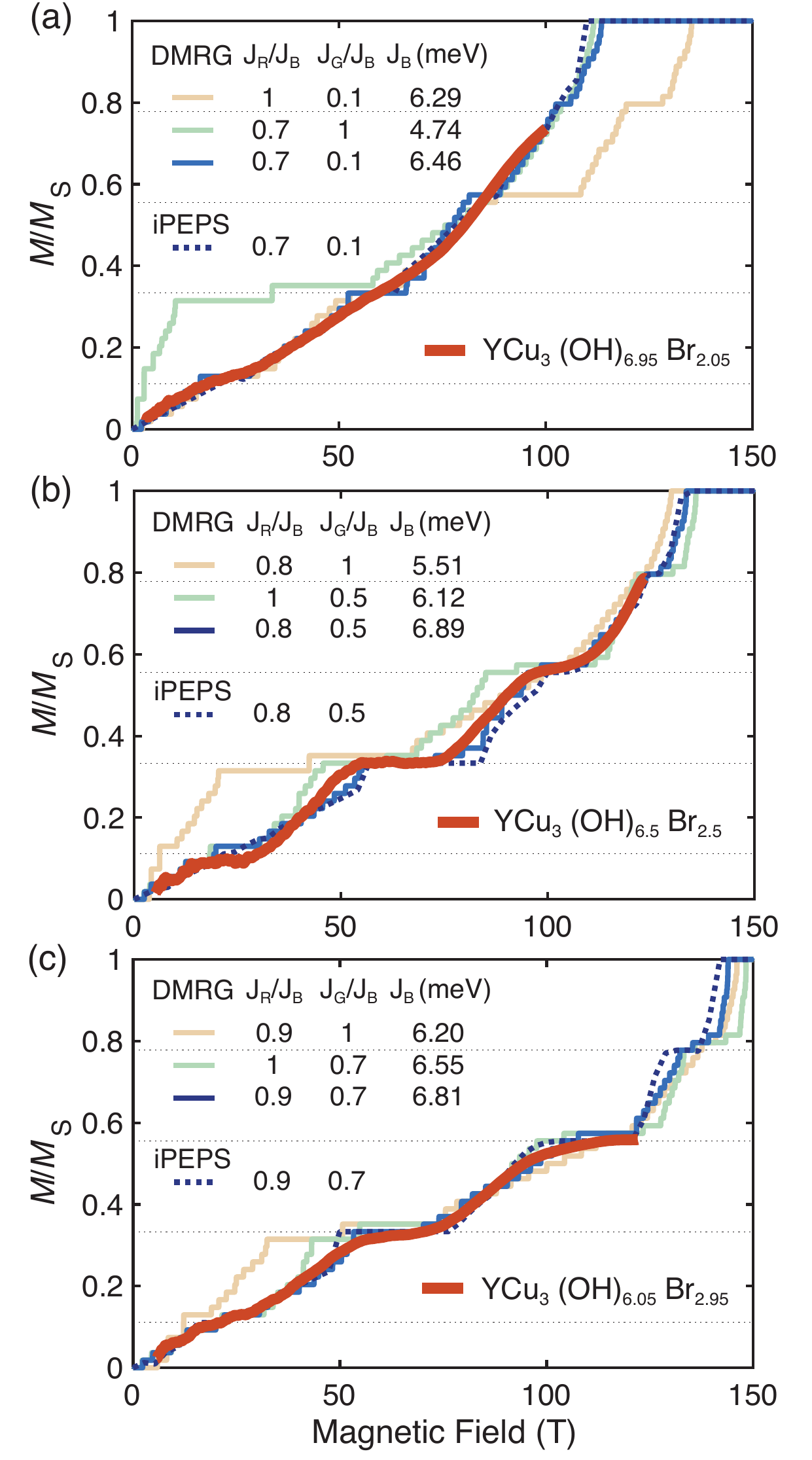}
     \caption{
     (a-c) Fitting results of the $M$-$H$ curves for YCu$_3$(OD)$_{7-x}$Br$_{2+x}$ materials 
     with different Br compositions. 
     The red solid lines represent the experimental data,
     while the solid blue and dashed blue curves correspond to the calculated results by DMRG and iPEPS with best-fitting parameters, respectively.
     The gray horizontal dotted lines indicate the expected $n/9$ ($n$=1, 3, 5, 7) magnetization plateaus.
     For comparison, more fittings obtained with different parameter sets are also shown in each panel.
     }
     \label{DMRGiPEPs}
    \end{center}
    \end{figure}

    \begin{figure}[b!]
    \begin{center}
    \includegraphics[width = 1\linewidth]{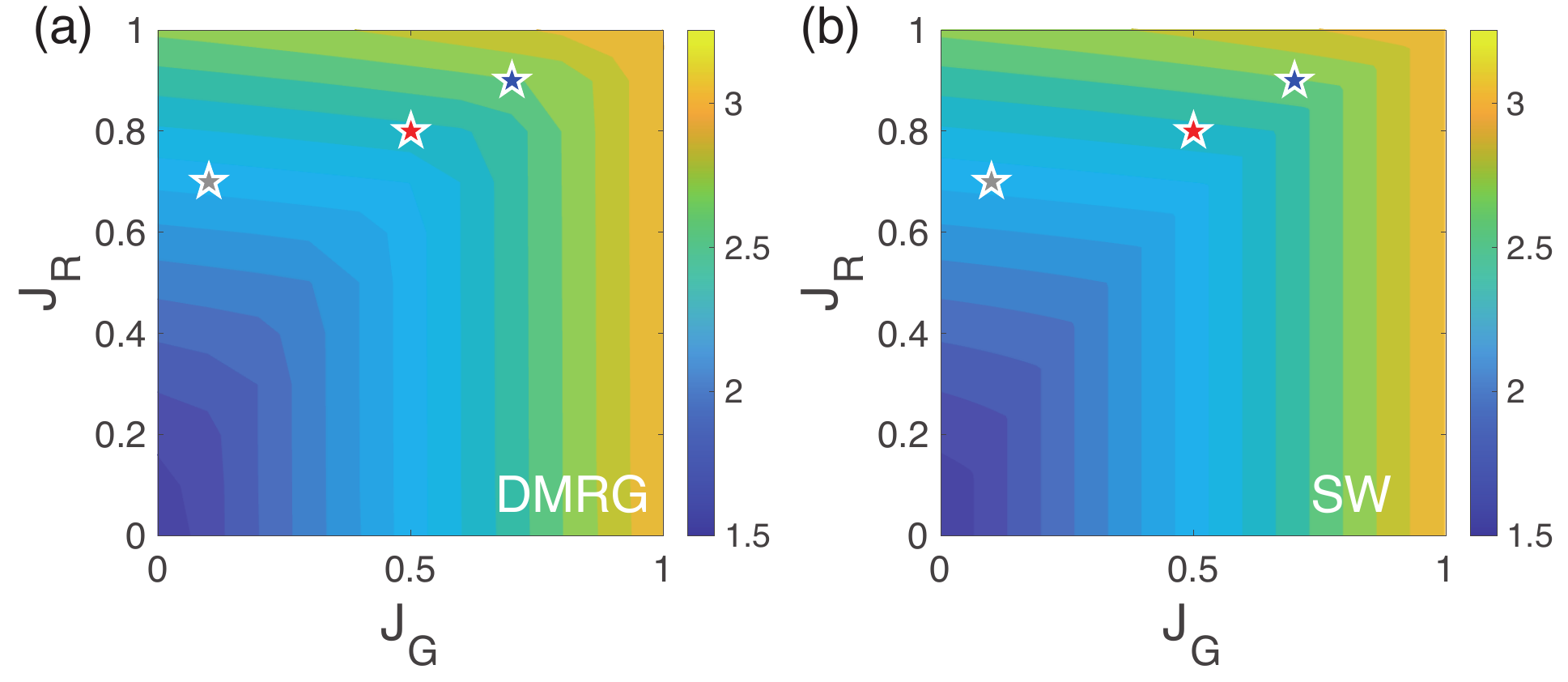}
     \caption{
     Contour plots of the calculated saturation field obtained by (a) DMRG and (b) spin wave (SW) methods, respectively. 
     The results are in natural unit with $J_{\rm B} = 1$.
     The best-fitting parameters for three materials are indicate by the stars, and the contour interval is set to $\Delta H =0.1$ across all panels.}
     \label{SW}
    \end{center}
    \end{figure}

For materials YCu$_3$(OD)$_{7-x}$Br$_{2+x}$ with different compositions, 
we visualize the loss function values in the parameter space using a color map through DMRG and iPEPS methods, 
as shown in the Fig.~\ref{DMRGfittings}(a-c) and Fig.~\ref{DMRGfittings}(d-f), respectively.
The overall good-fitting distributions from DMRG and iPEPS are consistent. 
We observe that under different parameters, 
the good-fitting regime (the dark blue area) gradually shifts toward the isotropic Heisenberg limit
as the Br composition increases from $x=0.05$ to $x=0.95$,
and the best fitted parameter sets (marked by the yellow stars) are shown in Table.~\ref{tab:params} of the main text.
The corresponding $M$-$H$ curves compared with experimental data are shown in Fig.~\ref{DMRGiPEPs}(a-c),
where we also present more magnetization curves under different parameters for comparison.\\

\noindent\textit{Comparison with spin wave results at saturation fields---}
As an additional check on the reliability of our DMRG calculations at saturation fields, 
we further perform spin wave (SW) analysis starting from the ferromagnetic state.
The DMRG results and the exact saturation fields are shown in Fig.~\ref{SW}(a) and Fig.~\ref{SW}(b), respectively.
Within the explored parameter range, 
the value of $J_{\rm R}$ is the primary determinant of the saturation field. 
It is noteworthy that the saturation fields obtained from our SW results for each parameter set are in excellent agreement with those derived from the DMRG calculations, where the minor difference arises from the grid resolution used in sampling the parameter space for each method. 
Therefore, it implies that finite-size effects and truncation errors in DMRG calculations are negligible under high fields.



%

\end{document}